ISME2020-2114

# Metal nanopowder production by cryogenic mechanical collision and pulsed electric-discharge methods

Arash Rahmati[1], Amir Abdullah[2], Kaveh Esmailnia[3]

[1]Department of Mechanical Engineering, Amirkabir University of Technology, Tehran; a_rahmati@aut.ac.ir
[2]Department of Mechanical Engineering, Amirkabir University of Technology, Tehran; amirah@aut.ac.ir
[3]Department of Mechanical Engineering, Amirkabir University of Technology, Tehran; kavehesmailnia@gmail.com

**Abstract**
With the introduction of micro and nanosensors and devices, the need for pure-metal and ceramic powders have risen. Spherical pure metal powders and their alloys are vital raw materials for near-net-shape fabrication via powder-metallurgy manufacturing routes as well as feed stocks for powder injection molding, and additive manufacturing. Employing atomization processes to achieve spherical powders dates back to 1980s and different attitudes to maintain a plasma current have been developed including gas atomization, plasma atomization, plasma-rotating-electrode atomization, and freefall atomization. Facilities for employing the aforementioned methods have always been expensive. This paper proposes two new processes by which pure spherical powder is achievable with relatively low costs. The first method proposed will deal with attrition of coarser particles in cryogenic milieu via coinciding jets (dubbed as cryogenic-coinciding jets or CCJ), while the second proposed method concerns melting and evaporation of metals caused by emitted heat from a plasma channel made by electric field breakdown at micrometer separations of metallic electrodes (entitled as electrode-plasma-atomization method or EPA). The CCJ method may face challenges such as cold welding of particles due to high speed collision and inefficiency of process because of presence of high plastic deformation and coldworking, whereas the EPA method demands only the use of electrically conductive materials as electrodes. Titanium powder with particle average sizes of 62nm and also 300-400nm was achieved via the EPA method. Studies also suggest that the CCJ method is expected to produce Titanium nanopowder with less than %0.05wt contamination.

## 1. Introduction

The field of nanotechnologies is advancing rapidly and is expected to impact virtually every factor of global industry and society [1]. Numerous solution techniques have been developed to synthesis metal, alloy, and ceramic micro/nanopowders with improved properties to achieve merit characteristics [2-5]. Nanoscale materials are identified with typical sizes on the order of 10-100nm and exhibit unique chemical, physical, optical, magnetic and mechanical properties [6]. One of the factors that can be tailored to obtain appropriate properties is the particles size in powders. As a particle diameter minifies, the surface area decreases by the factor of two, whereas the volume diminishes with the factor of three in relation to its diameter, therefore, the surface-aria-to-volume ratio increases, invoking brand-new characteristics, that are not present in neither the bulk material nor larger sized particles; characteristics such as explosive and catalytic behaviors. For instance, gold nanoparticles are characterized by a prominent optical resonance in visible range, which is sensitive to environmental changes, size, and shape of the particles as well as to local optical interactions in resonant systems [7].

### 1.1. Synthesis methods of Titanium powder

In the past decades many efforts have been devoted to achieve fine Titanium powders with the most purity possible, but little have been successful [8].

Generally, powder production methods fall into two main categories; a) top-down methods, b) bottom-up methods. In a top-down method, the precursor of the process consists mainly of either larger material particles or bulk materials, like different variations of milling and atomization, to name but a few; in contrast, in bottom-up methods the powder is usually formed by chemical reactions, precursors of which are molecules and atoms (e.g. sol-gel process, gas condensation synthesis, chemical precipitation, etc.) [9].

In this section, the main focus will be on PRE and cryogenic ball milling processes because of their similarities with the implemented methods.

Plasma rotating electrode (PRE) process is a refinement process developed by Nuclear Metals Inc. in the 1960s [10], in which the metal electrode is melted by the arc from a tungsten-tipped cathode with 60-90mm diameter and rotational speeds in the order of 3,000-15,000 rpm, making the liquid melt spun-off from the electrode surface due to centrifugal forces, leaving droplets in the chamber. Rapid cooling solidifies droplets in the air and forms a spherical pure metal powder. S. Abkowitz reported the production of 150μm diameter spherical Titanium alloy (Ti-7Al-2Nb-1Ta) powder using the aforementioned method. Nevertheless, discrete tungsten particles were found in Ti64 hot-isostatic-pressed powder, which is detrimental to the fatigue properties. Replacing the heat source with a transferred arc plasma torch would avoid Tungsten inclusion and using He in the plasma gun improves heat transfer properties and arc characteristics [11]. Fig. 1 shows the schematics of plasma rotating electrode process.

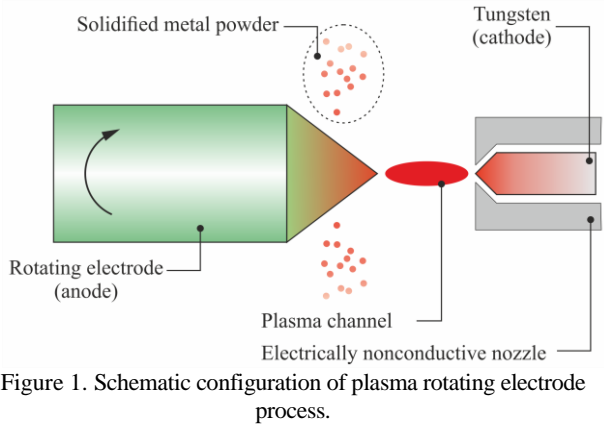

Figure 1. Schematic configuration of plasma rotating electrode process.

The dependence of the mean particle size ($d_{50}$) on the rotation speed of the electrode (S) and the electrode diameter ($D_e$) is expressed by the following equation [11], where K is the material constant determined by surface tension and density of the material.

$$d_{50} = \frac{K}{S\sqrt{D_e}} \quad (1)$$

The properties of bulk materials can be influenced by mechanical milling. The term "milling" itself refers to the reduction of particle sizes by mechanical forces. Cryogenic ball milling is a mechanical attrition technique in which powders are milled in a slurry formed with milling balls and a cryogenic liquid. The top-down approach of ball milling attracted attention because of its potential scalability. The mechanism of grain size refinement in a ball mill involves repeated welding and fracturing of powder particles. If the speed of welding surpasses the fracture rate, the mean size of powder particles increases, thus cooling down the milling chamber can not only defer welding but also decrease the fracture plastic deformation and energy required, which ultimately boosts the efficiency and the required time for the process. There is a vast variety of ball milling apparatuses and Fig 2. represents a simple common laboratory ball mill [12-14].

## 2. Proposed methods for producing metal powder

In this section, the description of proposed methods, as well as the calculation for the required equipment, will follow.

### 2.1. Pulsed-electric-discharge method

This approach employs the break-down of a high voltage electric field as a high energy, small-area-focused heat source. Two electrodes of the desired powder material are placed adjacently with a sub-micron, gap-controlled distance. An electric circuit (shown in Fig. 3.) charges the electrodes with a high voltage making a spark and a plasma channel, all happening in a high frequency (more than 100 kHz).

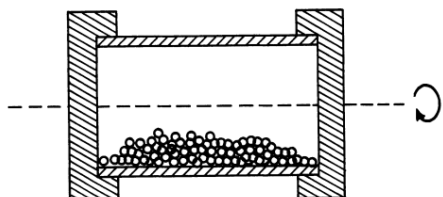

Figure 2. Schematics of a laboratory ball mill [13].

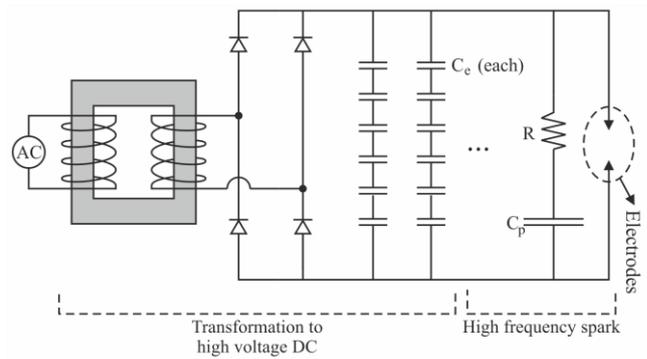

Figure 3. Concept configuration of electric circuit for EPA method.

The generated heat causes local evaporation of electrodes in the spark chamber and then the metal gas solidifies quickly, leaving fine, spherical, pure metal powders in the chamber. The schematic configuration of the apparatus is available in Fig. 3 and the schematics of the electric circuit are represented in Fig. 4. Using a slurry in the chamber would not only boost production rate and cooling rate but also could it reduce required voltage and current, making the process expeditious and efficient. The required voltage for electric field breakdown at atmospheric pressure against micrometer separation is plotted in Fig. 5. Considering available literature, it seems that the required voltage is independent of electrode material [15-17]. The calculations of required equipment follow:

Considering Fig 5., using a 500V transformer would appear to be acceptable and the breakdown seems to happen at 300V; Thus, Vs = 500V and Vb = 300V. The time duration of each spark will be derived 2μs from the equation given below:

$$T = R \times C_p \times \ln(\frac{V_s}{V_s - V_b}) \quad (2)$$

There are different values for polyester capacitor capacity to choose from and by choosing 220nF the ideal work exerted by each spark can be calculated:

$$W = \frac{1}{2} C_p V_b^2 \quad (3)$$

Thus, the spark work is calculated 0.0099j. This amount of work may seem little but the fact that it is done in just one individual spark within 2μs should be beard in mind.

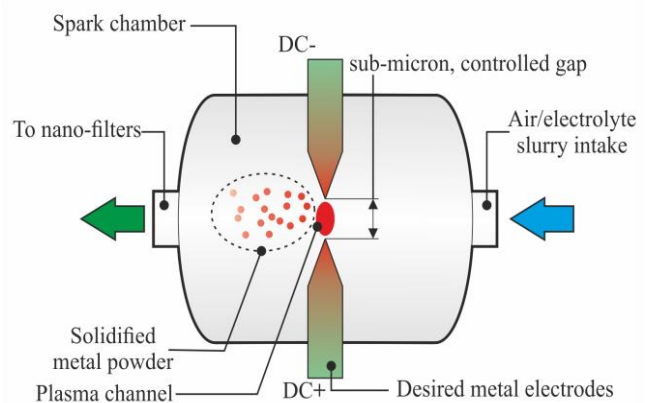

Figure 4. Breakdown voltage in air versus gap distance.

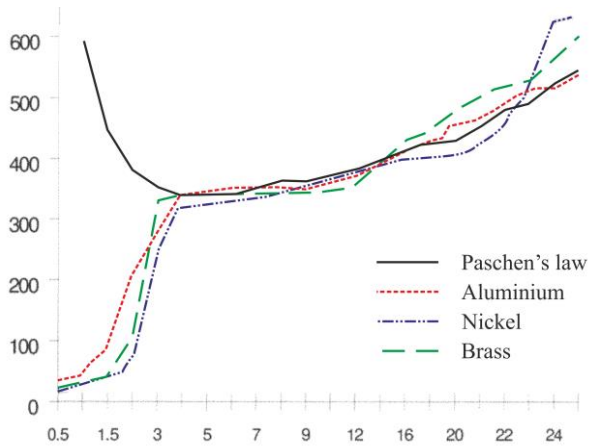

Figure 5. Breakdown voltage in air versus gap distance [14].

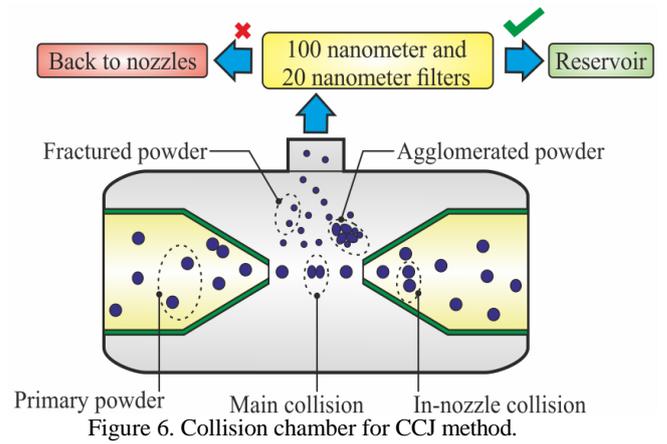

Figure 6. Collision chamber for CCJ method.

In ideal condition the total work exerted will be used to evaporate a globe of titanium without any loss:

$$Q = W \tag{4}$$

$$Q = m \times [C_{solid}(T_{fusion} - T_{ambient}) + L_f + \tag{5}$$
$$C_{liquid}(T_{evaporation} - T_{fusion}) + L_v]$$

$$m = \frac{4}{3}\pi r^3 \rho \tag{6}$$

By substituting eq. 5 in eq. 4 and solving eq.3 one can conclude that a titanium sphere with a diameter of 5030nm can be evaporated in each spark.

Assuming that spark time is equal to the required time for charging the capacitor, every second, n sparks occur:

$$n = \frac{1}{2T} \tag{7}$$

By substituting T from eq.1 in eq. 6 the number of sparks will be 248 thousand times per second, and the work done will be 2455j, which is enough to evaporate a sphere with the diameter of no less than 31.66**μ**m/s.

Decreasing spark-capacitor capacity would diminish the time duration of each spark, practically expected to make finer powder.

## 2. 2. Cryogenic-coinciding-jets method

In this method, the concepts of cryomilling are assimilated and used to come-up with an efficient approach in which powders are made to collide with a high speed, at a very low temperature (around -196˚C) in which most metallic materials are known to lose ductility and consume less energy to fracture [18], however, this does not apply to all materials and there is little data available for some materials. This simple method employs relatively low-cost facilities, mostly available in every workshop including an air compressor, some pipes, at least nanometric filters, and two hard-material nozzles. The powders are cooled and conveyed by a cryogenic gas and then collided in collision chamber. The powder will be made to collide until a proper size is achieved (see Fig. 6).

## 3. Experimental setup and procedure

To approve the applicability of EPA method, experiments were conducted via an EDM-204 apparatus from Akram Co. with titanium electrodes, oxygen gas flow and a fixed source voltage of 250V. The test was conducted within 3.5 hours and by swapping the electrodes poles every 30 minutes to prevent local agglomeration of powders and sticking of powders to electrodes. The configuration is presented in Fig 7.

In CCJ method, the fracture of particles is caused by mechanical collision and local evaporation as a result of heat generation from the impact, analogous to high-energy cryogenic ball milling. For approving the applicability of CCJ method, considering physical similarities of the mentioned method with cryogenic ball milling, the conclusions of other researchers would suffice. Kozlik et al. [12] used a cryogenic ball milling apparatus to synthesis $TiO_2$ in liquid nitrogen and liquid argon, with WC-Co and stainless-steel balls. They deliberately added a little amount of contamination to some experiments in order to prevent cold welding of the particles.

## 4. Results and discussion

Dynamic light scattering was performed on EDM-produced powder and the polydispersity index was found to be 1.6 and two peaks were observed in the curve as reported in Fig. 8. %28 of particles had a size average of 62nm and the remaining 415nm. This shows the effect improper filtering which allows agglomerates to pass into the reservoir. It also proves the ability of proposed, EPA method to achieve nanometric powder.

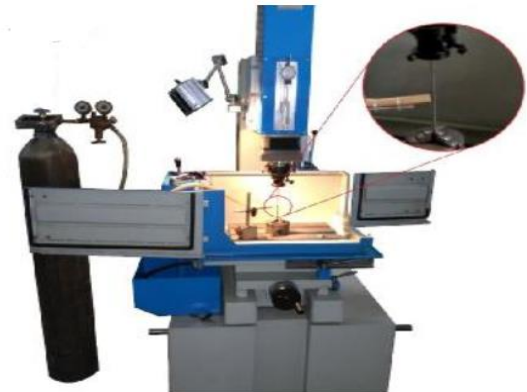

Figure 7. The test apparatus for assessing the applicability of EPA method.

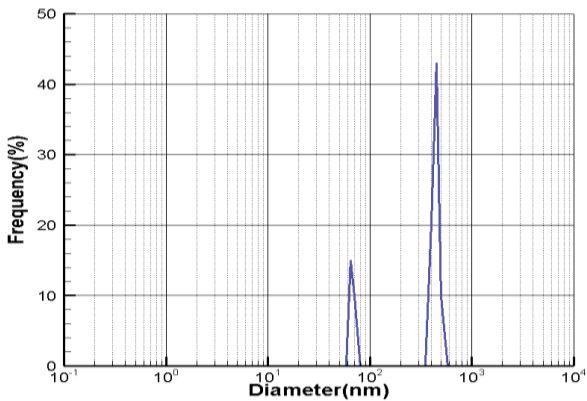

Figure 8. Dynamic light scattering results for EDM-produced powder.

Kozlik et al. reported that Ti powder particles do not significantly grow smaller by cryomilling unless when %4-8wt contamination of stearic acid was added. They were able to achieve sub-micron powder within only 3.25hrs with the grain size of approximately 100nm and high fraction of high angle boundaries. Their results approve that using CCJ method, one is able to obtain Titanium sub-micron powder with less than %0.05wt. hydrogen contamination and very little amount of nitrogen, carbon or oxygen contamination.

## 5. Conclusion

Two elementally-different methods and calculations for required equipment were provided in this article as well as a description for similar approaches to produce sub-micron metal powder. Applicability of proposed methods were put to test using EDM apparatus that achieved nanometric titanium powder was achieved and by cryomilling that resulted ultra-fine-grained titanium.

**Acknowledgement**
The authors would like to acknowledge Akram Co. staff for their support and cooperation and are thankful for the assistance of Miss Mahtab Rahmati and Mr. Pooria Chamani, master-of-science students in mechanical engineering, whose help and partnership were immensely of benefit and appreciated.

**Nomenclature**

$C_p$   Polyester-capacitor capacity

$Q$   Heat received by electrode

$R$   Electric resistance

$T$   Spark-on time

$V_b$   Breakdown voltage

$V_s$   Source voltage

$W$   Spark work

## References

[1] Document No. ISO/TR ISO/TR 12885: 2008 (E). Nanotechnologies—health and safety practices in occupational settings relevant to nanotechnologies.

[2] Sankar S, Sharma SK, An N, Lee H, Kim DY, Im YB, Cho YD, Ganesh RS, Ponnusamy S, Raji P, Purohit LP, 2016. "Photocatalytic properties of Mn-doped NiO spherical nanoparticles synthesized from sol-gel method". Optik. Nov 1;127(22):10727-34.

[3] Qin H, Guo W, Liu J, Xiao H, 2019. "Size-controlled synthesis of spherical ZrO2 nanoparticles by reverse micelles-mediated sol-gel process". Journal of the European Ceramic Society. Oct 1;39(13):3821-9.

[4] Kumar N, Biswas K, Gupta RK, 2016. "Green synthesis of Ag nanoparticles in large quantity by cryomilling". RSC advances, 6(112):111380-8.

[5] Kuiry, S.C., Megen, Ed., Patil, S.D., Deshpande, S.A., Seal, S., 2005. "Solution-Based Chemical Synthesis of Boehmite Nanofibers and Alumina Nanorods". J. Phys. Chem. B. 109 3868–3872

[6] Panáček A, Kvitek L, Prucek R, Kolář M, Večeřová R, Pizúrová N, Sharma VK, Nevěčná TJ, Zbořil R., 2006. "Silver colloid nanoparticles: synthesis, characterization, and their antibacterial activity". The Journal of Physical Chemistry B. Aug 24;110(33):16248-53.

[7] Sharma S, Reddy AV, Jayarambabu N, Kumar NV, Saineetha A, Rao KV, Kailasa S., 2019. "Synthesis and characterization of Titanium dioxide nanopowder for various energy and environmental applications". Materials Today: Proceedings. Nov 30.

[8] Behera PS, Sarkar R, Bhattacharyya S., 2016. "Nano alumina: a review of the powder synthesis method". Interceram-International Ceramic Review. Apr 1;65(1-2):10-6.

[9] S.A. Miller and P.R. Roberts, 1990. Powder Metal Technologies and Applications (Materials Park: ASM International), 1st ed., Vol. 7. ASM Handbook, pp. 97–101.

[10] Sun P, Fang ZZ, Zhang Y, Xia Y., 2017. "Review of the methods for production of spherical Ti and Ti alloy powder". Jom. Oct 1;69(10):1853-60.

[11] Kozlík J, Stráský J, Harcuba P, Ibragimov I, Chráska T, Janeček M., 2018. "Cryogenic milling of Titanium powder". Metals. Jan;8(1):31.

[12] Ganguli D, Chatterjee M., 1997 Jan. Ceramic powder preparation: a handbook. Boston: Kluwer Academic Publishers.

[13] Witkin DB, Lavernia EJ., 2006. "Synthesis and mechanical behavior of nanostructured materials via cryomilling". Progress in Materials Science. Jan 1;51(1):1-60.

[14] Torres JM, Dhariwal RS., 1999. "Electric field breakdown at micrometre separations". Nanotechnology. Mar;10(1):102.

[15] Ono T, Sim DY, Esashi M., 2000. "Micro-discharge and electric breakdown in a micro-gap". Journal of Micromechanics and Microengineering. Sep;10(3):445.

[16] Chen CH, Yeh JA, Wang PJ., 2006. "Electrical breakdown phenomena for devices with micron separations". Journal of Micromechanics and Microengineering. May 30;16(7):1366.

[17] Beer FP, Johnston ER, DeWolf JT., 1999. Mechanics of materials, 5th SI Edition. Stress;1(10):1-2.